\documentstyle[preprint,prb,aps]{revtex}
\begin{document}
\draft
\title{Bosonization  and phase Diagram of the\\ one-dimensional 
$t-J$ model}

\author{Guang-Hong Chen and Yong-Shi Wu}
\address{Department of Physics, University of Utah, 
Salt Lake City, Utah  84112}

\maketitle
\begin{abstract}
\baselineskip=0.90cm
We present an analytic study of the phase 
diagram of the one-dimensional $t-J$ model 
and a couple of its cousins. To deal with 
the interactions induced by the no double 
occupancy constraints, we introduce a 
deformation of the Hubbard operators. When 
the deformation parameter $\Delta$ is small, 
the induced interactions are softened, 
accessible by perturbation theory. We combine 
bososnization with renormalization group 
techniques to map out the phase diagram of the 
system. We argue that when $\Delta\to 1$, there 
is no essential change in the phase diagram. 
Comparison with the existing results in the 
literature obtained by other methods justifies 
our deformation approach.  
\end{abstract}


\newpage

\section{Introduction}
\subsection{Historical review}
Since the discoveries of quantum Hall effects and 
high $T_c$ oxides in 1980's, the strongly correlated 
systems have been of great interests both theoretically 
and experimentally. As far as the high $T_c$ problem is 
concerned, the $t-J$ model is believed to be the 
appropriate starting model Hamiltonian, because it 
captures the essence of the interplay between charge 
and spin degrees of freedom in superconducting $Cu$ 
oxides \cite{anderson,fczhang}. Although high-$T_c$ 
cuprates are (at least) two-dimensional systems, it 
is very interesting to study its one-dimensional (1D) 
counterpart. As argued by Anderson\cite{anderson}, 
two-dimensional strongly correlated systems may share 
some properties of 1D systems. In addition, the physical 
understanding of the 1D systems is also extremely helpful 
for the study of the ladder systems, which have been 
realized experimentally and have attracted a lot of attention 
in recent years\cite{dagotto,experiments,lin-balents-fisher}. 

In the 1D systems, the phase space of particle 
scattering is highly restricted. The occurrence of a 
single scattering event will spread quickly among all 
other particles, which invalidates the concept of 
individual excitations. Consequently, we are often 
confronted with correlated collective excitations. 
On the other hand, in some cases we can benefit from 
such phase space restriction. Namely, the many-particle 
scattering matrix could be nicely decomposed into the 
product of two-particle ones which satisfy the so-called 
Yang-Baxter integrable conditions\cite{yang}. This 
property provides us with the possibility to exactly 
solve some 1D models, e.g., the Hubbard model\cite{lieb-wu}, 
the Heisenberg model\cite{bethe,yang-yang}, and the 
supersymmetric $t-J$ model\cite{supertj}. The exact 
solutions in turn provide us with powerful guidelines 
to develop and to justify certain approximate schemes 
for other problems. 

In some sense, the 1D $t-J$ model could be viewed as 
a descendant of the Hubbard model in large on-site 
repulsion limit.  Namely, the strong coupling limit 
of the Hubbard model can be mapped into the weak 
coupling limit of the $t-J$ model. Naively, one may 
speculate that the integrability of the 1D Hubbard 
model would be inherited by the $t-J$ model in the 
whole parameter space. Unfortunately, this 
speculation is not correct: The $t-J$ model is only 
integrable at two special points in parameter space. 
The reason for this difference is that, in contrast 
to other 1D integrable models, the Hilbert space 
of the $t-J$ model is highly constrained: Double 
occupancy of any site is completely excluded. 
Furthermore, the integrable points of the $t-J$ 
model are located in the strong coupling regime 
($|J|=2t$ in our convention below), not in the 
weak coupling limit. Therefore, the integrability 
of the supersymmetric $t-J$ model is not simply 
inherited from the Hubbard model. Rather it is  
better to be viewed as a separate miracle of the
interacting 1D many-particle system. Since the 
1D $t-J$ model can not be exactly solved at a 
generic point in parameter space, the analytical 
studies of the $t-J$ model have been a painstaking 
task even in the 1D case. 

To illustrate the points more clearly, let us take 
a close look at the $t-J$ model. The model 
delineates the behavior of hard core fermions on 
a discrete lattice, and the dynamics is given by 
the model Hamiltonian
\begin{equation}
\label{tjhamilton}
H_{tJ}=-t{\cal P}\sum_{j,\sigma}
(c^{\dag}_{j,\sigma}c_{j+1,\sigma}+h.c.){\cal P}
+J\sum_{j}{\bf S}_j\cdot{\bf S}_{j+1}. 
\end{equation} 
Here ${\cal P}$ is the projection operator 
that prohibits double occupancy of any site, 
$\sigma$ and $\bar{\sigma}\equiv-\sigma$ the 
spin orientations (with $\sigma$=1 for $\uparrow$, 
and -1 for $\downarrow$); $t$ is the hopping 
amplitude and $J$ the anti-ferromagnetic ($J>0$) 
or ferromagnetic ($J<0$) coupling. Due to the 
aforementioned constraints, at each site the 
states ${|a>}$ can only be one of the following 
three possible states: with $a=\uparrow,\downarrow$, 
and $a=0$ (empty). This Hilbert space is neither 
fermionic nor bosonic. One can check that the 
projection operators $\chi^{ab}=|a><b|$ close, 
under commutation and anti-commutation, to form 
a semi-simple supersymmetric Lie algebra, the 
$Spl(1,2)$ given by the relations\cite{wiegmann}
\begin{equation}
\label{spl12}
\{\chi^{ab}_i,\chi^{cd}_j\}_{\pm}
=\delta_{ij}(\chi^{ad}_i\delta^{bc}
\pm\chi^{bc}_i\delta^{ad}), 
\end{equation}
where $\chi^{\sigma 0}_i$ and $\chi^{0\sigma}_i$ 
are fermionic operators that, respectively, create 
and annihilate a single electron. The bosonic 
operator $\chi^{\sigma\sigma^{\prime}}$ are 
identified as the generators of the group $SU(2)$. 
Using these operators, the $t-J$ model can be 
neatly written as
\begin{equation}
\label{tj2}
H_{tJ}=-t\sum_{j,\sigma=\uparrow,\downarrow}
(\chi^{0\sigma}_j\chi^{\sigma 0}_{j+1}+h.c.)
+J\sum_{j,\sigma,\sigma^{\prime}}
\chi^{\sigma\sigma^{\prime}}_j
\chi^{\sigma^{\prime}\sigma}_{j+1},
\end{equation}
in terms of the bilinears in the generators of 
$Spl(1,2)$. But the price we have to pay is to 
introduce both fermionic and bosonic operators 
simultaneously. As a consequence, it is difficult 
to make a simple, controlled approximation in 
this representation. To overcome the difficulties
associated with the no-double-occupancy 
constraints, the slave boson and slave fermion 
method\cite{coleman,read} and, more recently, 
the supersymmetric Hubbard operator 
method\cite{tsvelik} have been invented to treat 
the $t-J$ model, with the hope of the mean field 
ground state being relevant to the high-$T_c$ 
problem. However, after more than one decade 
effort, it seems that a reliable ground state 
is still elusive. 

\subsection{Deformed Hubbard operators}

The seminal work by Jordan and Wigner\cite{jordan} 
and, later, by Lieb, Schultz, and 
Mattis\cite{mattis} provides an alternative 
idea to handle the above hybridized situation in 
statistics: Namely the spin operators are uniformly 
expressed in terms of fermionic operators, though 
the spin systems are neither bosonic nor fermionic 
ones. In the same spirit, one can also rewrite
the $t-J$ model in terms of fermions exclusively.

In addition to rewriting the magnetic 
interactions using the fermionc realization 
of the spin operators
\begin{equation}
\label{spinrealization} 
\vec{S}_j=\frac{1}{2}c^{\dag}_{j\alpha}
\vec{\sigma}_{\alpha\beta}c_{j\beta},
\end{equation}
one may introduce as well the Hubbard 
operators\cite{hubbard} 
\begin{equation}
\label{hubbard}
\bar{c}^{\dag}_{j\sigma}=
c^{\dag}_{j\sigma}(1-n_{j\bar{\sigma}})
\end{equation}
and rewrite the hopping terms in terms of 
them. These operators also realize the 
constraints that exclude double occupancy 
on each lattice site. In this way, one 
gets a formulation of the $t-J$ model 
completely in terms of fermionic operators.

However, one immediately sees that the old 
hopping terms will induce extra four-fermion 
and six-fermion interactions. These interactions 
are "hard" ones, in the sense that their 
strengths are exactly the same as the hopping 
amplitude $t$. This fact defies the attempts 
to treat the additional four- and six-fermion
terms perturbatively. Therefore, at first 
glance, it looks silly to adopt this strategy 
to solve the $t-J$ model unless new techniques 
can be invented to make the induced interactions 
tractable. 

In the present paper we propose a technique 
that allows us to deal with these induced four- 
and six-fermion interactions. The key point is 
to use the idea of "{\it adiabatic continuity}" 
{\it to soften} the above-mentioned interactions 
induced by the no-double-occupancy constraints. 
Namely, we propose to introduce the following 
{\it deformed} Hubbard operators
\begin{equation}
\label{deformed}
\bar{c}^{\dag}_{j\sigma}=c^{\dag}_{j\sigma}
(1-\Delta n_{j\bar{\sigma}})
\end{equation}
with a deformation parameter $0<\Delta\le 1$. 
When $\Delta$ approaches to unity, we recover 
the genuine Hubbard operators 
(\ref{hubbard}). For $0< \Delta <1$, there is
a non-zero probability to allow leakage into 
states with double occupancy. With these deformed 
Hubbard operators (\ref{deformed}) replacing the 
genuine Hubbard operators (\ref{hubbard}) in the 
hopping terms, we get a deformation of the 
original $t-J$ model. The deformed model has the 
advantage that for small $\Delta$, the induced 
four- and six-fermion interactions are no longer 
``hard''. This is because these interactions 
have strengths proportional to the deformation 
parameter $\Delta$ and, therefore, is tractable
in the sense of perturbation theory when 
$\Delta$ is small. 

Though small values of $\Delta$ may not be 
``physical'', after extracting possible 
structures in the phase diagram for small 
$\Delta$, we analytically continue our results 
back to $\Delta=1$. The fundamental assumption 
underlying this continuation is the {\it 
adiabatic continuity}, namely that when the 
Hamiltonian of the model is adiabatically 
changed with $\Delta$ varying from a small positive 
value to unity, there is no essential, 
qualitative change in the phase diagram of the 
model, though various phase boundaries in 
parameter space may undergo a continuous 
deformation. Historically, our idea of 
considering a deformed model is parallel 
to the ideas that underlie the replica 
method in treating disordered system, or the 
large $N$ expansion in field theory. Actually, 
even in the field of 1D exactly solvable 
models one can find a precedent: Yang and 
Yang\cite{yang-yang} proposed the $XXZ$ model 
as a deformation of the $XXX$ model, i.e. the 
spin-$\frac{1}{2}$ 1D Heisenberg model, and used 
it to justify the Bethe Ansatz method in the 
latter by first studying the large anisotropic 
limit and then continuing back to the isotropic 
limit. In this paper we will first discuss some 
simpler cases and give arguments to justify 
the adiabatic continuity assumption together 
with our deformed Hubbard operators.  

Of course, practically the justification may
depend on how we treat the deformed model, 
which is a fully fermionized model containing 
four- and six-fermion interactions. In the present 
paper, we are going to combine the bosization 
method and perturbative renormalization group 
(RG) techniques to deal with the deformed $t-J$ 
model. Namely we first bosonize the deformed 
model, and then use the RG flows to map out 
the phase diagram of the bosonized model. We
will argue that the phase diagram obtained in 
this way does not change in an essential way, 
when the deformation parameter $\Delta$ varies 
from a small positive value to unity.   

For convenience, we will start with a 
simplified model. Namely, we will first 
consider a model in which the magnetic 
spin-spin interactions are Ising-like, i.e. 
of the form $J_{z}\sum_{j}S^{z}_jS^{z}_{j+1}$.
This model, together with the usual hopping 
term, we call the $t-J_z$ model. Then with 
a bit more complication, we would like to 
modify the isotropic magnetic interactions 
in eq. (\ref{tjhamilton}) to anisotropic 
$XXZ$-type interactions:
\begin{equation}
\label{tjmodify}
H=-t{\cal P}\sum_{j,\sigma}
(c^{\dag}_{j\sigma}c_{j+1,\sigma}+h.c.){\cal P}
+J_{\perp}\sum_{j}(S^{x}_jS^{x}_{j+1}
+S^{y}_jS^{y}_{j+1}) 
+J_{z}\sum_{j}S^{z}_jS^{z}_{j+1}.
\end{equation}
This model we call as the $t-J_{\perp}-J_z$ model.  
The phase diagram of the $SU(2)$ invariant $t-J$ 
model can be obtained in the double limit with 
$J_{\perp}\to J_z$ (the isotropic limit) and 
with the deformation parameter $\Delta\to 1$ 
(the physical limit with no double occupancy).

The paper is organized as follows: In 
section II, we discuss the phase structure 
of the extremely anisotropic limit of the 
$t-J_{\perp}-J_z$ model, namely, the $t-J_z$ 
model. The convention of our bosonization 
scheme is also presented in detail in this 
section. Then the discussions of the 
phase diagram  for the 1D $t-J_{\perp}-J_z$ 
model are presented in the section III. In 
section IV, we compare our results with other 
work. The discussions and conclusions are 
summarized in the section V.

\section{An extremely anisotropic limit: 
the $\hbox{t}$-$J_{z}$ model}

\subsection{The model}

The 1D $t-J_z$ model represents a strongly 
anisotropic limit of the $SU(2)$ $t-J$ model, 
in which only has the Ising part of the 
magnetic interactions been included. Without 
hopping, this simplification is significant
for understanding purely magnetic interactions. 
However, with hopping the model is more 
interesting in that it has incorporated 
the interplay between hopping and the 
exchange interactions, which makes the physics 
of the model highly non-trivial. Therefore,
the model has recently attracted a lot of 
interests\cite{los}. It is known from the 
numerical studies that the low-energy physics 
in both the $t-J_z$ and $t-J$ models\cite{dagotta2} 
shares some common features even in two 
dimensions. In the real world, the possible 
origin of exchange anisotropy is the spin-orbital 
coupling\cite{bonsteel}. In the extremely 
anisotropic limit, the Hamiltonian (for the 
$t-J_z$ model) reads
\begin{eqnarray}
\label{tJz}
{\cal H}_{tJ_z}&=&-t\sum_{j\sigma}
(\bar{c}^{\dag}_{j\sigma}\bar{c}_{j+1,\sigma}
+H.c.)+J_z\sum_{j}S^{z}_jS^{z}_{j+1}\\ \nonumber
&=&H_0(t)+U(J_z).
\end{eqnarray}
Following Eq. (\ref{spinrealization}), we use
the representation of $S^{z}_j$ given by
\begin{equation}
\label{sz}
S^{z}_j=\frac{1}{2}(n_{j\uparrow}-n_{j\downarrow}).
\end{equation} 
Note the appearance of the Hubbard operators 
(\ref{hubbard}) in the hopping terms. It is the 
presence of the second term in eq. (\ref{hubbard}) 
that realizes the no double occupancy constraints,
As a consequence, the term $H_0(t)$ is no longer 
a simple hopping of fermions: More interaction 
terms with four or six fermions are induced, with
strengths of the same order of magnitude as the
hopping amplitude $t$. How to deal with these 
interaction terms is an important issue. 

To reduce the strengths of the interaction terms
induced by the no double occupancy constraints,
we propose to deform the model Hamiltonian (\ref{tJz}) 
by replacing the Hubbard operators with the deformed 
Hubbard operators (\ref{deformed}), resulting in
\begin{equation}
H_0(t)=H_h+H_1+H_2+H_3,
\end{equation}
The Hamiltonians $H_i$($i=h,1,2,3$), in terms 
of the genuine fermion operators $c_{j\sigma}$ 
and $c^{\dag}_{j\sigma}$, are given by
\begin{equation}
\label{Hh}
H_h=-t\sum_{j\sigma}
(c^{\dag}_{j\sigma}c_{j+1,\sigma}
+H.c.),
\end{equation}
which represents the genuine hopping term, and
\begin{eqnarray}
\label{hi}
H_1&=& t\Delta\sum_{j\sigma}(c^{\dag}_{j\sigma}c_{j+1,
\sigma}n_{j+1,\bar{\sigma}}
+H.c.); \\
H_2&=& t\Delta\sum_{j\sigma}(c^{\dag}_{j\sigma}c_{j+1,
\sigma}n_{j\bar{\sigma}}+H.c.); 
 \\
H_3&=& -t\Delta^2\sum_{j\sigma}(c^{\dag}_{j\sigma}
c_{j+1,\sigma}n_{j\bar{\sigma}}n_{j+1,\bar{\sigma}}+H.c.). 
\end{eqnarray}

Here $H_1$ and $H_2$ are the induced four fermion 
repulsive interaction to prevent double occupancy 
of the same lattice site, and the $H_3$ term is 
attractive, representing the effects 
from the six fermion interactions that compensate 
to the excessive  repulsion  in $H_1$ and $H_2$. 
It is easy to  see that now in the deformed model, all the induced 
terms $H_1$, $H_2$ and $H_3$ are proportional to 
the deformation parameter $\Delta$ in Eq. (\ref{deformed}). 
If $\Delta$ is small, the induced interactions are 
"softened", becoming tractable in perturbation theory. 
In the limit of $\Delta\to 1$, the total effects of the 
three terms precisely prevent double occupancy for each 
lattice site. 

By using Eq. (\ref{sz}), the exchange 
term $U(J_z)$ is given by
\begin{equation}
\label{ujz}
U(J_z)=\frac{J_z}{4}\sum_j(n_{j\uparrow}-n_{j\downarrow})
(n_{j+1,\uparrow}-n_{j+1,\downarrow}).
\end{equation}
In this way, we rewrite the $t-J_z$ model in terms of 
fermion creation and annihilation operators exclusively.
To look for the low energy effective Hamiltonian, we
perform the standard procedure to bosonize
the $t-J_z$ model in the next subsection.

\subsection{Bosonization}

The hopping term is easily diagonalized by 
Fourier transform, the energy spectrum 
is given by
\begin{equation}
\label{barespectrum}
\varepsilon(k)=-2t\cos(ka),
\end{equation}
where $a$ is lattice spacing. In the ground
state, all the states with momentum lower than
the Fermi momentum $k_F$ are filled. For a 
generic filling factor $\nu=N/M$ with $N$ the 
particle number and $M$ the number of lattice 
sites, the Fermi momentum is 
\begin{equation}
\label{fermimomentum}
k_Fa=\frac{\pi}{2}\nu.
\end{equation}
To get the low energy effective action for the 
excitations, we only need to focus on momenta 
close to $\pm k_F$ and linearize the spectrum as
\begin{equation}
\label{linearspectrum}
\varepsilon(\pm k_F+q)=\pm v_Fq-2t\cos(k_Fa),
\end{equation}
where the Fermi velocity is given by 
$v_F=2ta\sin(k_Fa)$. The second term is a constant 
and can be shifted away by redefining the energy 
zero point. We will drop it throughout the rest 
of the paper. 

In one dimension, the definition of exchange 
statistics is ambiguous, since the no double 
occupancy condition excludes the possibility
to physically exchange spatial position of 
two particles. This makes the statistics of 
fermionic particles lose its absolute meaning 
and make an alternative description in terms 
of bosons possible. This situation is quite 
different from that of the three dimensional
case, where the exchange statistics 
of particles has an absolute meaning. In two 
dimensions, the definition of particle statistics 
only marginally makes sense and we can transmute 
the statistics arbitrarily by attaching the 
Chern-Simons flux to particles (the composite 
of particle and flux is dubbed as anyon
\cite{frank,wu}). The statistics transmutation 
procedure in one dimension is called bosonization; 
it has been widely used in exploring the physics 
in one dimensional systems\cite{haldane,schulz-delft}.

In practice, it is convenient to discuss the 
bosonization in real space. To do so, we expand 
the lattice fermion in terms of continuum fields
\begin{eqnarray}
\label{continuumexpansion}
c_{j\sigma}&=&\sqrt{a}[\psi_{R\sigma}(x)
e^{ik_Fx}+\psi_{L\sigma}(x)e^{-ik_Fx}] \\ 
c^{\dag}_{j\sigma}&=&\sqrt{a}
[\psi^{\dag}_{R\sigma}(x)e^{-ik_Fx}
+\psi^{\dag}_{L\sigma}(x)e^{ik_Fx}], 
\end{eqnarray}
where $x=ja$ is used. After linearization and 
dropping fast varying terms, we get the low energy 
effective Hamiltonian for the $H_h$ term as
\begin{equation}
\label{lowHh}
H_h=\int dx {\cal H}_{h}(x)=-v_F\sum_{\sigma}
\int dx[\psi^{\dag}_{R\sigma}i\partial_x\psi_{R\sigma}
-\psi^{\dag}_{L\sigma}i\partial_x\psi_{L\sigma}],
\end{equation}
describing a one-dimensional relativistic Dirac 
particle in continuum.

In the following, we use the bosonization 
rule to bosonize the $t-J_z$ model:
\begin{equation}
\label{bosonization}
\psi_{P}(x)=\frac{\eta_{P}}{\sqrt{2\pi a}}
e^{i\phi_{P}(x)},
\end{equation}
where $\eta_{P}$ ($P=R/L=+/-,\sigma,...$) are 
the Klein factors to maintain the anti-commuting 
relations between particles on different sites. 
We can also realize the Klein factor as Majarona 
fermions which satisfy the following anti-commuting 
relations
\begin{equation}
\label{majarona}
\{\eta_r,\eta_{s}\}=2\delta_{rs}.
\end{equation}
The introduction of $1/\sqrt{2\pi a}$ in Eq. 
(\ref{bosonization}) maintains the correct
dimension for the field $\psi_{P}(x)$ which 
has dimension $[length]^{-1/2}$ (see Eq. 
(\ref{continuumexpansion})). The $\phi$ fields 
are angular variables and thus dimensionless. 
To get the correct anticommutation for fermionic 
fields, we also require the bosonic fields 
$\phi_{P}(x)$ satisfy the following commutation 
relations
\begin{equation}
\label{commutation}
[\phi_{P\sigma},\phi_{P\sigma^{\prime}}]= 
iP\pi\delta_{\sigma\sigma^{\prime}}
\varepsilon(x-x^{\prime});
\end{equation}
\begin{equation}
[\phi_{R\sigma},\phi_{L\sigma^{\prime}}]
=i\pi\delta_{\sigma\sigma^{\prime}},
\end{equation}
where $\varepsilon(x)$ is the Heaviside step 
function.

Using these bosonization rules, we get the 
bosonic description for the hopping term 
(\ref{lowHh}) as
\begin{equation}
\label{bosonicHh}
H_{h}=\frac{v_F}{4\pi}\sum_{\sigma}
\int dx[(\partial_x\phi_{R\sigma})^2
+(\partial_x\phi_{L\sigma})^2].
\end{equation}
For later convenience, let us introduce a pair 
of conjugate non-chiral bosonic fields for each 
species:
\begin{eqnarray}
\label{nonchiral}
\phi_{\sigma}&\equiv &\phi_{R\sigma}+\phi_{L\sigma}; \\
\theta_{\sigma}&\equiv&\phi_{R\sigma}-\phi_{L\sigma},
\end{eqnarray}
which satisfy
\begin{equation}
\label{relnonchiral}
[\phi_{\sigma}(x),\theta_{\sigma^{\prime}}(x^{\prime})]
=-i4\pi\delta_{\sigma\sigma^{\prime}}
\varepsilon(x-x^{\prime}).
\end{equation}
To organize the spin and charge modes more elegantly, 
we introduce new pairs of dual fields as follows:
\begin{eqnarray}
\label{spincharge}
\phi_c(x)&=&\frac{1}{\sqrt{2}}(\phi_{\uparrow}
+\phi_{\downarrow});\hspace{1.5cm}\phi_s(x)
=\frac{1}{\sqrt{2}}(\phi_{\uparrow}-\phi_{\downarrow}),\\
\theta_c(x)&=&\frac{1}{\sqrt{2}}(\theta_{\uparrow}
+\theta_{\downarrow});
\hspace{1.6cm}
\theta_s(x)=\frac{1}{\sqrt{2}}(\theta_{\uparrow}
-\theta_{\downarrow}).
\end{eqnarray} 
Here the introduction of numerical factor $1/\sqrt{2}$ 
is to maintain the commutation relation in Eq. 
(\ref{relnonchiral}). The subscript ``s'' means the 
spin mode and ``c'' the charge mode. Using these 
spin-charge separated modes, the bosonic Hamiltonian 
(\ref{bosonicHh}) can be cast into the following form
\begin{equation}
\label{scHh}
H_h=\frac{v_F}{8\pi}\int dx[(\partial_x\phi_c)^2
+(\partial_x\phi_s)^2 +(\partial_x\theta_c)^2
+(\partial_x\theta_s)^2].
\end{equation}
 
To bosonize the induced interaction terms $H_{1,2}$, 
we note that, roughly speaking, both the $H_1$ and 
$H_2$ terms are of the type of Hubbard-like on-site 
interactions in the continuum limit and, therefore, 
provide interactions to renormalize the charge/spin 
velocity and the controlling parameters (i.e. $K_{c/s}$; 
see below) and the cosine term in the spin sector. 
By taking microscopic details of these two terms into 
account, we get an extra numerical factor 
$\cos(k_Fa)=\cos(\pi\nu/2)$. Both terms have the same 
bosonized form.  Namely, the bosonized form of 
$H_1+H_2$ is
\begin{eqnarray}
\label{H12}
H_1+H_2&=&2H_1 \\ \nonumber
&=&\frac{t\Delta a\cos(k_Fa)}{\pi^2}
\int dx[(\partial_x\theta_c)^2 
-(\partial_x\theta_s)^2]+\frac{4t\Delta{\cal P}
\cos(k_Fa)}{\pi^2a}\int dx \cos(\sqrt{2}\theta_s) 
\\ \nonumber
&=&\frac{v_F\Delta\cot(k_Fa)}{2\pi^2}
\int dx[(\partial_x\theta_c)^2 
-(\partial_x\theta_s)^2]
+\frac{2v_F\Delta{\cal P}\cot(k_Fa)}{\pi^2a^2}
\int dx \cos(\sqrt{2}\theta_s),
\end{eqnarray}
where ${\cal P}=\eta_{R\uparrow}\eta_{L\uparrow}
\eta_{R\downarrow}\eta_{L\downarrow}$, since 
${\cal P}^2=1$, we get ${\cal P}=\pm 1$; in the 
following, we will take ${\cal P}=+1$.

Now we come to discuss the six-fermion term $H_3$ 
in the continuum limit; after a straightforward 
but tedious calculation we get
\begin{eqnarray}
\label{h3}
H_3=&-&\frac{\sqrt{2}t\Delta^2a^2\cos(k_Fa)}{4\pi^3}
\int dx\partial_x\theta_c(x+a)[(\partial_x\theta_c)^2
-(\partial_x\theta_s)^2] \\ \nonumber
&-&\frac{t\Delta^2\cos(k_Fa)}{\sqrt{2}\pi^3}
\int dx\partial_x\theta_c(x+a)\cos(\sqrt{2}\theta_s) 
\\ \nonumber
&+&\frac{t\Delta^2\sin(k_Fa)}{\sqrt{2}\pi^3}
\int dx\partial_x\theta_s(x+a)
\sin(\sqrt{2}\theta_s).
\end{eqnarray}
To get a sensible continuum limit, we have to take 
$a\rightarrow 0$, but keep $ta$ finite consistently. 
An elegant way to accomplish this is to use the 
following operator product expansion (OPE):
\begin{eqnarray}
\label{OPE}
\partial_{z_1}\theta_c(z_1)\partial_{z_2}\theta_c(z_2)
&\sim&-\frac{1}{(z_1-z_2)^2}; 
\\ \nonumber
\partial_{z}\theta_s(z)\sin[\sqrt{2}\theta_s(0)]&
\sim&-\frac{\sqrt{2}}{z}
\cos[\sqrt{2}\theta_s(0)],
\end{eqnarray}
with all other OPEs being regular. We finally get 
the continuum limit of the six-fermion term to be
\begin{equation}
\label{finalH3}
H_3=-\frac{v_F\Delta^2}{2\pi^3a^2}
\int dx\cos(\sqrt{2}\theta_s).
\end{equation}

The last thing in bosonizing the $t-J_z$ model is 
to bosonize the magnetic interaction $U(J_z)$ in 
Eq. (\ref{ujz}). We decompose it into the 
following combinations:
\begin{eqnarray}
\label{Jz}
U(J_z)&=&\frac{J_z}{4}\sum_{j}(n_{j\uparrow}
-n_{j\downarrow})(n_{j+1\uparrow}-n_{j+1\downarrow}) 
\\ \nonumber
&=&\frac{J_z}{4}\sum_{j}(n_{j\uparrow}n_{j+1\uparrow}
+n_{j\downarrow}n_{j+1\downarrow})-\frac{J_z}{4}
\sum_{j}(n_{j\uparrow}n_{j+1\downarrow}
+n_{j\downarrow}n_{j+1\uparrow}).
\end{eqnarray}
The terms in the first bracket are Coulomb interactions 
between the electrons on different sites; in the 
continuum limit we get its bosonized form as
\begin{equation}
\label{bosonicCoulomb}
\frac{J_z}{4}\sum_{j}(n_{j\uparrow}n_{j+1\uparrow}
+n_{j\downarrow}n_{j+1\downarrow})=\frac{J_za}{16\pi^2}
\int dx[(\partial_x\theta_c)^2+(\partial_x\theta_s)^2].
\end{equation} 
The terms in the second bracket of eq. (\ref{Jz}) are 
the Hubbard-like interactions in the continuum limit; 
the bosonization procedure gives
\begin{equation}
\label{hubbardlike}
\frac{J_z}{4}
\sum_{j}(n_{j\uparrow}n_{j+1\downarrow}
+n_{j\downarrow}n_{j+1\uparrow})=\frac{J_za}{16\pi^2}
\int dx[(\partial_x\theta_c)^2-(\partial_x\theta_s)^2]
+\frac{J_z\cos(2k_Fa)}{4\pi^2a}\int dx\cos(\sqrt{2}
\theta_s).
\end{equation}
Combining Eqs. (\ref{bosonicCoulomb}), (\ref{hubbardlike})
with Eq. (\ref{Jz}), we find that the terms involving the 
charge variable $\theta_c$ exactly cancel and we get the 
bosonized form 
of $U(J_z)$ as
\begin{equation}
\label{Ujz}
U(J_z)=\frac{J_za}{8\pi^2}
\int dx (\partial_x\theta_s)^2
-\frac{J_z\cos(2k_Fa)}{4\pi^2a}
\int dx \cos(\sqrt{2}\theta_s).
\end{equation}
It is worth noting that the absence of charge variables 
in the $U(J_z)$ term is natural, since we are dealing 
with pure magnetic interactions. 

Therefore, after collecting all the results, the low energy 
effective Hamiltonian for the 
bosonized form of the $t-J_z$ model is nicely written as
\begin{equation}
\label{bosonHtjz}
{\cal H}_{tJ_z}=\int dx (H_c+H_s),
\end{equation}
where the Hamiltonian for the spin sector ($H_s$) 
and the charge sector ($H_c$) are given by
\begin{equation}
\label{charesector}
H_c=\frac{v_c}{2}[K_c\Pi^2_c+\frac{1}{K_c}
(\partial_x\theta_c)^2],
\end{equation}
\begin{equation}
\label{spinsector}
H_s=\frac{v_s}{2}[K_s\Pi^2_s+\frac{1}{ K_s}
(\partial_x\theta_s)^2]+\frac{g_{\theta}}
{8\pi^2a^2}\cos(\sqrt{2}\theta_s),
\end{equation}
where $\Pi_{c}=\frac{1}{4\pi}\partial_x\phi_c$ and 
$\Pi_{s}=-\frac{1}{4\pi}\partial_x\phi_s$ are 
the conjugate momenta for the charge field $\theta_c$ 
and spin field $\theta_s$ respectively. The effective 
coupling constant $g_{\theta s}$ is given by
\begin{equation}
\label{effective1}
g_{\theta}=v_F
\biggr[8\Delta\cot(\frac{\pi}{2}\nu)
+\frac{J_za}{v_F}\cos(\pi\nu)
-\frac{2\Delta^2}{\pi}\biggl].
\end{equation} 
The velocities $v_{c/s}$ are renormalized by 
magnetic interactions and the interactions 
induced by the no double occupancy conditions: 
\begin{eqnarray}
\label{vcvs}
v_c&=&v_F\sqrt{1+\frac{4\Delta}{\pi}
\cot(\frac{\pi}{2}\nu)}; \\
v_s&=&v_F\sqrt{1+\frac{J_za}{\pi v_F}
-\frac{4\Delta}{\pi}\cot(\frac{\pi}{2}\nu)}.
\end{eqnarray}
The controlling parameters $K_{c/s}$ are 
given by
\begin{eqnarray}
\label{vsks}
K_c&=&\frac{4\pi}{\sqrt{1+\frac{4\Delta}{\pi}
\cot(\frac{\pi}{2}\nu)}};\\ 
K_s&=&\frac{4\pi}{\sqrt{1+\frac{J_za}{\pi v_F}
-\frac{4\Delta}{\pi}\cot(\frac{\pi}{2}\nu)}}. 
\end{eqnarray}
In passing, we would like to stress that the 
above results are derived for small $\Delta$ 
and $J_za$. However, the general result of 
a renomalization of $v_{c/s}$ and $K_{c/s}$, 
but with no other changes, is expected
to be valid more generally\cite{gang,sachdev}. 
In other words, the functional forms of the 
low energy effective Hamiltonians $H_{c/s}$, 
being basically dictated by the symmetry 
requirements, survive even if the interactions 
are strong, while the above values of $v_{c/s}$ 
and $K_{c/s}$ are not universal. Therefore, we 
conclude that if we adiabatically continue the 
value of $\Delta$ to unity, the low energy 
effective Hamiltonian of the $t-J_z$ model 
should be of the same form as the above 
charge and spin Hamiltonians $H_{c/s}$, 
with the renormalized values of $v_{c/s}$ 
and $K_{c/s}$ not restricted to those given 
by Eqs. (\ref{vcvs}) and (\ref{vsks}). 

\subsection{The phase diagram}

Now we are in the position to discuss the possible 
phase diagram for the $t-J_z$ model. For convenience, 
we only discuss the anti-ferromagnetic case, namely, 
we assume $J_z>0$.

At first, we notice that the spin and charge degrees
of freedom are well separated just like what happened 
in other 1D interacting models. However, from 
the expression for the controlling parameters 
$K_{c/s}$, we have already seen the interesting 
interplay between hopping and magnetic interactions. 
The phase diagram is determined by the competition 
of above two energy scales( $t$ and $J_z$). This is 
quite different from the case of the Hubbard model 
or the $XXZ$ model where the controlling parameter 
is only determined by the interaction strength. However, 
the charge sector is massless, described by a quadratic 
Hamiltonian with no mass term. This means that charge 
excitations are gapless and the charged sector of the 
system is metallic. In contrast, the knowledge on
the fate of the spin sector needs more work. The 
situation is similar to that of the Hubbard model.

The fate of the spin sector is determined by the 
well-studied sine-Gordon Hamiltonian. In the 
spin sector we have the following renormalization
group equations (RGE)\cite{sachdev,gang}:
\begin{eqnarray}
\label{rg}
\frac{dg_{\theta}}{dl}&=&
(2-\frac{K_s}{2\pi})g_{\theta}; \\ 
\frac{dK_s}{dl}&=&-\delta g^2_{\theta}.
\end{eqnarray}
where $\delta$ is a positive, regularization 
dependent parameter. With these two RG equations 
in hand, we can readily analyze the phase 
diagram for the spin sector.

(i) When $K_s>4\pi$ and $|g_{\theta}|\le 
(\frac{K_s}{4\pi}-1)/\sqrt{\delta}$, the spin 
sector flows to the fixed point line:
\begin{eqnarray}
\label{LLfixed}
g^{*}_{\theta}&=&0, \\ 
K_s&>&4\pi.
\end{eqnarray}
Thus we get the Luttinger liquid behavior  
for the spin sector. Following the Balents-Fisher's 
notation\cite{lin-balents-fisher}, we say that the 
system is in the $C1S1$ phase; here more generally
a $CmSn$ phase means a phase with $m$ massless 
charge modes and $n$ massless spin modes 
respectively.

(ii) When parameters $K_s$ and $g_{\theta}$ satisfy 
one of the following conditions:
\begin{equation}
\label{spin-peierls1}
K_s\le 4\pi, \hspace{1.0cm} g_{\theta}>0;
\end{equation}
or
\begin{equation}
\label{spin-peierls2}
K_s>4\pi, \hspace{1.0cm}  
g_{\theta}<(\frac{K_s}{4\pi}-1)/\sqrt{\delta},
\end{equation}
then the RGE flows toward $g_{\theta}=+\infty$. 
In this case, the behavior of the system is overwhelmingly 
determined by the minima  of the cosine term. 
For $g_{\theta}>0$, these minima are given by
\begin{equation}
\label{minima1}
\theta_s=\sqrt{2}(n+\frac{1}{2})\pi,
\end{equation}
but due to the angular nature of the variable $\theta_s$, 
we can have only two distinct ground states, distinguished 
by the even and odd values of $n$. This state is identified 
to be Peierls ordering of spin degrees of freedom. Due to 
quantum tunneling, degeneracy of the ground state is removed. 
Consequently, the excitations above either ground state
 are gapful. The dominant contributions to the 
mass gap come from the topological soliton excitations in 
the dilute gas approximation of solitons and anti-solitons. 
Therefore, in this phase, the spin sector is gapful, and we
classify the phase of the system as a $C1S0$ phase.

(iii) In contrast to the case (ii), if the parameter $K_s$ 
and $g_{\theta}$ satisfy one of the following two conditions:
\begin{equation}
\label{neel1}
K_s\le 4\pi, \hspace{1.0cm} g_{\theta}<0;
\end{equation}
or
\begin{equation}
\label{neel2}
K_s>4\pi, \hspace{1.0cm}  
g_{\theta}<-(\frac{K_s}{4\pi}-1)/\sqrt{\delta},
\end{equation}
then the RGE flows toward $g_{\theta}=-\infty$. An 
argument similar to that in the case (ii) gives the 
ground states determined by
\begin{equation}
\label{minima2}
\theta_s=\sqrt{2}n\pi.
\end{equation}
In this state, we have a staggered expectation value
for the $z-$component of the spin. Therefore, 
the spin ordering is Neel-like.

In summary, we construct the phase diagram for the 
$t-J_z$ model in Figure 1. 

\section{The $\hbox{t}-J_{\perp}-J_z$ model}

In this section, we will discuss the modified 
version (\ref{tjmodify}) of the $t-J$ model. 
Again the change to make is in the magnetic 
interactions. In addition to the $U(J_z)$ term
discussed in the last section, we now add the 
$XY$ part, $U(J_{\perp})$, of the 
anti-ferromagenetic interactions: 
\begin{eqnarray}
\label{Uperp}
U(J_{\perp})&=&J_{\perp}
\sum_{j}(S^{x}_jS^{x}_{j+1}+S^{y}_jS^{y}_{j+1}) 
\\ \nonumber
&=&\frac{J_{\perp}}{2}\sum_{j}(c^{\dag}_{j\uparrow}
c_{j\downarrow}c^{\dag}_{j+1,\uparrow}c_{j+1,\downarrow}
+c^{\dag}_{j\downarrow}c_{j\uparrow}
c^{\dag}_{j+1,\downarrow}c_{j+1,\uparrow}). 
\end{eqnarray}
Following the bosonization procedure presented in the 
preceding section, we get the bosonized form for 
$U(J_{\perp})$ as
\begin{equation}
\label{bosonicUperp}
U(J_{\perp})=\frac{J_{\perp}}{2\pi^2a}
\int dx\cos(\sqrt{2}\phi_s)\cos(\sqrt{2}\theta_s)
-\frac{J_{\perp}}{2\pi^2a}[1-\cos(2k_Fa)]
\int dx\cos(\sqrt{2}\phi_s).
\end{equation}

Therefore, for the modified $t-J$ model (\ref{tjmodify}), 
we have the bosonized low energy effective Hamiltonian
\begin{equation}
\label{bosonicTJ}
{\cal H}=\int dx (H_c+\tilde{H}_s),
\end{equation}
where the Hamiltonian of the charge sector, $H_c$, is 
still given by Eq. (\ref{charesector}), since the 
$XY$ part of magnetic interactions only changes 
spin dynamics. In contrast, due to the extra 
$U(J_{\perp})$, the Hamiltonian $H_s$ of the spin 
sector has been drastically modified to
\begin{eqnarray}
\label{hsTJ}
\tilde{H}_s&=&\frac{v_s}{2}[K_s\Pi^2_s
+\frac{1}{K_s}(\partial_x\theta_s)^2] \\ \nonumber
&+&\frac{g_{\theta}}{8\pi^2a^2}
\int dx\cos(\sqrt{2}\theta_s) -\frac{g_{\phi}}{8\pi^2a^2}
\int dx\cos(\sqrt{2}\phi_s) \\  \nonumber
&+&\frac{g_{\theta\phi}}{8\pi^2a^2}
\int dx\cos(\sqrt{2}\phi_s)\cos(\sqrt{2}\theta_s).
\end{eqnarray}
Compared to the $t-J_{z}$ model, the Hamiltonian $H_s$ 
now has two extra terms with the coupling constants 
$g_{\phi}$ and $g_{\theta\phi}$ respectively. The spin 
velocity ($v_s$) and controlling parameter ($K_s$) are 
still the same as those in Eqs. (\ref{vcvs}-\ref{vsks}). 
Using Eq. (\ref{bosonicUperp}), the coupling constants 
$g_{\phi}$ and $g_{\theta\phi}$ are determined to be
\begin{eqnarray}
\label{cons}
g_{\phi}&=&8J_{\perp}a\sin^2(\frac{\pi}{2}\nu); \\
g_{\phi\theta}&=&4J_{\perp}a.
\end{eqnarray}

Due to the appearance of the interaction term 
$g_{\theta\phi}$, which has a non-zero conformal 
spin, the dynamics for the spin sector becomes 
much more involved. When we use the scaling arguments 
to discuss the relevance of the interaction terms, 
we need to be more careful.  We'd better use  
the RG flow for the Hamiltonian (\ref{hsTJ}) 
to discuss the details of the spin dynamics. 
Fortunately, up to one loop level, the RGE for 
a Hamiltonian like (\ref{hsTJ}) have been studied 
thoroughly, though in quite different context
\cite{kus,yak,gang}. The resulting RGE for the 
double cosine term $g_{\theta\phi}$ is
\begin{equation}
\label{RGE}
\frac{dg_{\theta\phi}}{dl}
=2\biggr[1-\biggr(\frac{K_s}{4\pi}
+\frac{4\pi}{K_s}\biggl)\biggl]g_{\theta\phi}. 
\end{equation} 
Since we know 
\begin{equation}
\label{inequality}
\frac{K_s}{4\pi}+\frac{4\pi}{K_s}\ge 2,
\end{equation}
the double cosine term in the
Hamiltonian (\ref{hsTJ}) is always irrelevant. 
Of course, the action of RG will generate 
more terms, such as single cosine terms. 
However, the arguments of these single cosine 
terms are twice bigger and these terms are more 
irrelevant than the existing terms. Thus we 
can neglect them. This situation is quite 
different from that of the two coupled Luttinger 
liquid case\cite{kus,yak,gang}. Therefore, we 
only need to focus on the following effective 
Hamiltonian
\begin{eqnarray}
\label{effectivespin}
\tilde{H}_s&=&\frac{v_s}{2}[K_s\Pi^2_s
+\frac{1}{K_s}(\partial_x\theta_s)^2] 
+\frac{g_{\theta}}{8\pi^2a^2}
\int dx\cos(\sqrt{2}\theta_s) 
-\frac{g_{\phi}}{8\pi^2a^2}
\int dx\cos(\sqrt{2}\phi_s), \\ \nonumber
&=&\frac{v_s}{2}[\tilde{\Pi}^2_s
+(\partial_x\tilde{\theta}_s)^2] 
+\frac{g_{\theta}}{8\pi^2a^2}
\int dx\cos(\beta_s\tilde{\theta}_s) 
-\frac{g_{\phi}}{8\pi^2a^2}
\int dx\cos(\tilde{\beta}_s\tilde{\phi}_s),
\end{eqnarray}
where we have introduced $\tilde{\Pi}_s=\sqrt{K_s}\Pi_s$ 
and $\tilde{\theta}_s=\frac{\theta_s}{\sqrt{K_s}}$.
The definitions of $\beta_s$ and $\tilde{\beta}_s$ are 
\begin{equation}
\label{beta}
\beta_s=\sqrt{2K_s},\hspace{1.0cm} 
\tilde{\beta}_s=\frac{16\pi}{\sqrt{2K_s}}.
\end{equation}
It is now easy to observe that the low energy effective
Hamiltonian possesses the following duality property: 
Namely, the Hamiltonian (\ref{effectivespin}) is 
invariant under the following transformation
\begin{equation}
\label{duality}
\beta_s\longleftrightarrow
\tilde{\beta}_s, \hspace{1.0cm}
g_{\theta}\longleftrightarrow -g_{\phi}.
\end{equation}
Note that such a duality does not appear in the $t-J_z$
model or in the Hubbard model. But it is also interesting 
to note that it appeared in the 1D $XYZ$ Thirring 
model\cite{gang} and in the case of two coupled 
Luttinger liquids\cite{kus,yak,gang}. 

Compared with the sine-Gordon system, the symmetry 
of Eq. (\ref{effectivespin}) is discrete, while there 
is a hidden $U(1)$ symmetry in the sine-Gordon system,
which reflects the $U(1)$ invariance of its dual
fermionic model (the Massive Thirring model). 

It is also easy to get the scaling dimension for the
cosine terms of the field $\theta_s$ and its conjugate 
$\phi_s$ as
\begin{equation}
\label{doublescaling}
\Delta_{\theta}=\frac{K_s}{2\pi},\hspace{1.0cm} 
\Delta_{\phi}=\frac{8\pi}{K_s}.
\end{equation}
Therefore, one of the two cosine terms is always 
relevant, which is associated with the ordering 
of the $\theta$- or $\phi$-field. Let us discuss the 
following two different cases separately.

(i) When the scaling dimension  $\Delta_{\theta}<2$, 
the $\cos(\sqrt{2}\theta_s)$ term is relevant. This 
case is similar to the $t-J_z$ case, and the system 
eventually flows to the spin-Peierls phase for 
$g_{\theta}>0$ or the Ising-Neel order for 
$g_{\theta}<0$ respectively.

(ii) When the scaling dimension $\Delta_{\phi}<2$, 
the $\cos(\sqrt{2}\phi_s)$ term is relevant. In this 
case, since $g_{\phi}$ is always negative, therefore
the system flows toward the Ising-Neel phase only.

In summary, we see that the phase diagram in Fig. 1 can
only be partially accessed in the $t-J$ model.
The difference in the two cases reflects the fact 
that the duality transformation (\ref{duality}) 
can only be realized in part of the parameter space,
since the coupling constant $g_{\phi}$ is definitely 
negative, while the coupling constant $g_{\theta}$ 
can be either negative or positive, depending on the 
interplay between $t$ and $J_z$.

\section{Conclusions and discussions}

In this paper, the phase diagram of the most general
1D $t-J_{\perp}-J_z$ model is discussed based on 
bosonization and RGE. To make sense of the bosonization 
procedure for the interactions induced by no double
occupancy constraints, we have introduced deformed 
Hubbard operators (\ref{deformed}), which contain
a deformation parameter $\Delta$. While at $\Delta=1$
the no double occupancy constraints at each site are 
recovered, the case with a small positive $\Delta$
is accessible to perturbative RG analysis. Since the
basic structure of the bosonized low energy effective
Hamiltonian is argued to be determined only by the 
symmetry requirements, the bosonized form of the low 
energy effective Hamiltonian with a small deformation 
parameter $\Delta$ is expected to survive the limit
$\Delta\to 1$. However, we can not simply use the 
values of $v_{c/s}$ and $K_{c/s}$ to make precise 
predictions on the phase diagram, since these values 
are not reliable at $\Delta=1$. We should take the
strategy in which both $v_{c/s}$ and $K_{c/s}$ are 
considered as  phenomenological parameters. 

For the case with $J_z\gg J_{\perp}$, the model is 
reduced to the so-called $t-J_z$ model.  In this case, 
the spin sector can flow to three distinct phases: the
gapless phase, the spin-Peierls phase, and the Ising-Neel 
phase, depending on the range of the parameters,
meanwhile the charge dynamics remains always gapless. 
In the case with $J_z>J_{zc}$, where the $J_{zc}$ 
represents the value to make $g_{\theta}=0$ and $K_s<4\pi$, 
the system flows to the Ising-Neel ordering in spin 
dynamics. We identify this phase as the so-called phase 
separation (PS) phase. For the case with $J_z<J_{zc}$ 
and $K_s<4\pi$, the spin sector eventually flows to 
the spin-Peierls phase which is gapful. We can identify 
this phase as a superconducting phase (SC). Finally, for 
$K_s>4\pi$, the spin sector flows toward a gapless phase 
and thus the system flows toward the Tomonoga-Luttinger 
liquid phase. Such a phase is consistent with the phase 
diagram constructed by Los Alamos group in 
Ref.\onlinecite{los}, where the authors mapped the $t-J_z$  
model into the 1D $XXZ$ model and construct the phase 
diagram from the knowledge of exact solutions for the 1d 
$XXZ$ model. This consistency also helps us to justify our
proposal to use the deformed Hubbard operators and the 
continuation from the case of $\Delta\ll 1$ to the desired
case $\Delta=1$.

In the opposite limit, namely $J_{\perp}\gg J_z$, the 
modified $t-J$ model can be reduced the the $t-J_{\perp}$
model. 
In this case, we still have $g_{\theta}$ generally non-zero 
due to the no-double-occupancy induced interactions. 
Therefore, the phase diagram of the $t-J_{\perp}$ model 
is expected to be similar to the case of the most general 
$t-J_{\perp}-J_z$ model. Namely, the system is generically 
gapful in the spin sector and thus can not flow toward 
the Tomonoga-Luttinger phase. This result is a little bit 
different from the naive speculation that the $t-J_z$ 
model should be  basically similar to the $t-J$ model. From our study, 
we conclude that there are some delicate differences between 
the two cases, since the $XY$ part and the Ising part of the 
magnetic interactions play a different role in spin ordering.

Acknowledgments

This research was supported in part by the US 
National Science Foundation under Grants No. 
PHY-9970701.


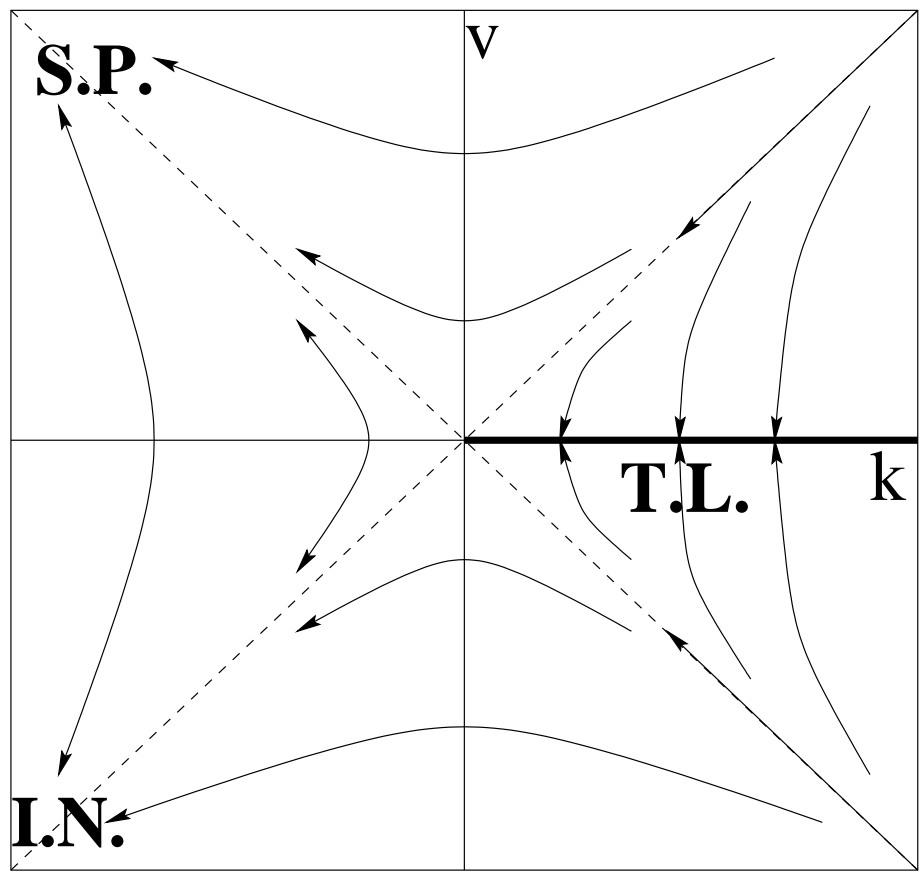
\begin{figure}
\caption{The schematic phase diagram. The RGE flow gives the
possible fate of $t-J_z$ model as the spin-Peiers phase (S.P.),
Ising-Neel pahse (I.N.), and Tomonoga-Luttinger phase (T.L.).}
\end{figure}
\end{document}